\documentclass[aps,prd,nofootinbib,twocolumn,superscriptaddress,showpacs]{revtex4}

\usepackage{graphicx}
\usepackage{amssymb}
\usepackage{amsmath}

\def\gsim{ \lower .75ex \hbox{$\sim$} \llap{\raise .27ex \hbox{$>$}} }
\def\lsim{ \lower .75ex\hbox{$\sim$} \llap{\raise .27ex \hbox{$<$}} }
\def\araa{Ann. Rev. Astron \& Astrophys.}
\def\aap{Astron \& Astrophys.}
\def\apjl{ApJL.}
\def\apj{ApJ.}
\def\aapr{Astron \& Astrophys. Rev.}
\begin{document}

\title{A New Limit on the Distances of Nuclei UHECRs Sources}

\author{Tsvi Piran}
\email{tsvi@phys.huji.ac.il}
\affiliation{Racah Institute of Physics, The Hebrew University, Jerusalem, Israel}

\begin{abstract}
Recent evidence from the Pierre Auger Observatory suggests a transition, at 5 EeV-10EeV  in the composition of Ultra High Energy Cosmic Rays (UHECRs), from protons to heavier nuclei such as iron. I consider here the implications of the heavier composition on  the sources of UHECRs. 
The smaller magnetic rigidity implies that  nuclei UHECRs are: (i) More easily accelerated (ii) Local, as they can diffuse only a few Mpc from their sources before disintegrating  (iii) Isotropic, because large deflections in the extra galactic and the galactic magnetic fields erase the  directional information.  
Uncertainties in the strength and structure of the the extra galactic magnetic field (EGMF)  makes it difficult to estimate the overall effects. However,  with  typical reasonable parameters  of a few nG and a coherence distance of a Mpc the distance  a nuclei UHECR above the GZK energy traverses before photodisintegrating  is only  a few Mpc.  In spite of the  significantly  weaker limits on the luminosity, Cen A is the only currently active potential source of nuclei UHECRs within this distance.  The large deflections erases the directional anisotropy expected from a single source.   
If indeed the composition of above-GZK-UHECRs is iron and if the EGMF is not too small then Cen A is the dominant  source of observed nuclei UHECRs above the GZK limit.

\end{abstract}

\pacs{96.50.S-, 13.85.Tp, 98.70.Sa,98,62En,98,70Sa}

\maketitle

\section{Introduction}

Recent observations of the Pierre Auger Observatory (PAO) show a transition in  $\langle {\rm X_{max}} \rangle $ and  
in $RMS ( {\rm X_{max}})$ between  5EeV and 10EeV \cite{Abraham10}. These transitions are interpreted as reflecting a transition in the composition of Ultra High energy Cosmic Rays (UHECR)  in this energy range from protons  (denoted hereafter as pUHECRs) to intermediate weight nuclei and in particular towards iron (denoted hereafter nUHECRs).  Like protons, iron nuclei also suffer a strong annihilation above the Greisen-Zatsepin-Kuzmin (GZK) \cite{Greisen66, ZatsepinKuzmin66} energy of $\sim 50$EeV. However, the larger charge and mass affect  both the acceleration and the propagation of 
nUHECRs leading to a drastically different view on their possible sources.


I begin by a brief review of the observations. 
(i) Both PAO \cite{AugerGZKsuppression08} and HiRes \cite{HiResGZKsuppression08} found a decline in the spectral slop of UHECRs above $\sim 50$EeV. This is roughly at the energy for which the Greisen-Zatsepin-Kuzmin (GZK) \cite{Greisen66, ZatsepinKuzmin66} suppresion due to the interaction of the UHECRs with the CMB is expected. The relevant interactions are photopion production for protons and photodisintegration for nuclei, however coincidentally, the threshold for both interaction is roughly at the same energy.  (ii) As already mentioned, the PAO reports a transition in composition from protons to intermediate mass nuclei, more specifically iron, between 
5EeV and 10EeV \cite{Abraham10}.  These results are 
consistent with a recent analysis of Yakustk data \cite{YakutskComposition08} but are 
not supported by HiRes that finds a protonic composition \cite{HiResComposition07}. 
(iii) The overall sky distribution of the arrival directions of
UHECRs below the GZK energy is  isotropic  
(see, e.g., \cite{NaganoWatson00,HiResIsotropyLow,AugerVarious09}). However, at higher energies, which are of more interest to us,   analyses of different experiments lead to somewhat  conflicting conclusions.
For example,  AGASA finds an excessive number of pairs and one
triplet in the arrival direction of CRs above 40EeV suggesting  small scale clustering \cite{AgasaIsotropy}. On the other hand, the HiRes stereo data are consistent with the hypothesis of null clustering
\cite{HiResIsotropy04} while the autocorrelation analysis of the
PAO data reported a weak excess of pairs for $E > 57$EeV \cite{AugerCorrelation08}. 
The PAO data shows a large scale isotropy \cite{AugerIsotropy08} however,  PAO \cite{AugerCorrelation08}  
found a correlation (within a  radius of 3.1$^o$) between  events above $57$EeV with  AGNs located closer than 75 Mpc in the V«eron-Cetty \& V«eron  catalog \cite{Veron}. 
The HiRes data, however, do not show  a correlation of the highest energy events with nearby AGNs \cite{HiresIsotropy08}, but there have been controvercial claims of correlation of the HiRes data with distant BL Lac objects \cite{BLLAC04}. Finally, the PAO also finds for $E>57$EeV  a "hot spot" in the direction of Cen A
\cite{AugerCenA}. Different authors make different cuts and attribute different number of UHECRs and a different statistical significance to this "hot spot".

I consider here the implications of an iron composition of the highest energy CRs on the acceleration (\S \ref{acceleration}) and on the propagation  (\S \ref{propagation}) of UHECRs. I examine the implications  to relevant sources in \S \ref{sources}. 

\section{Acceleration}
\label{acceleration}

The smaller magnetic rigidity  of the iron leads to  several  important differences on the nature of possible sources of nUHECRs as compared with sources of pUHECRs. First,  it eases the strict acceleration constraint\cite{Hillas84}:
\begin{equation}
R B >   10^{16} ~({\rm cm~ G })~ E_{20} \Gamma^{1} \beta^{-1} Z_{26}^{-1} ,
\label{Hillas}
\end{equation}
where $\Gamma$ and $\beta$ are the Lorentz factor and velocity of the source. Generally $Q_{x} \equiv Q/10^x$ and $E_{20}\equiv E/10^{20}$eV  but for the electric charge  $Z_{26} \equiv Z/26$, in units of the electron's charge. 
This criterion sets a limit on the Poynting flux luminosity of the source (see e.g. \cite{Waxman95}):
\begin{equation}
L > 1.5  \times 10^{42} ~ ({\rm erg/s})~ (\Gamma^2/\beta) E^2 _{20} Z^{-2}_{26} \ . 
\label{luminosity}
\end{equation}
A comparison of synchrotron losses with the  acceleration rate limits  the source's magnetic field:
\begin{equation}
B <  3~ {\rm G }~ \Gamma^2 E_{20}^{-2} (Z_{26}/A_{56})^{-4} Z_{26} \  , 
\label{Bfield}
\end{equation}
where $A_{56}$ is the atomic weight in units of the proton's mass. An additional condition on the magnetic field arises for nuclei. The energy of a synchrotron photon generated by the gyro motion of the nucleolus shouldn't be large enough to destroy the nucleolus:
\begin{equation}
\label{Bsynch}
B < 0.3 ~{\rm G}~ Z_{26} A^{-3}_{56} (h \nu_{synch}/ 5 {\rm MeV}).
\end{equation}
Both limits on the magnetic field in the source (Eqs. \ref{Bfield} and \ref{Bsynch}) are easily satisfied for AGNs.
A related condition arises on the strength of the radiation field at photon energies of a few MeV within the acceleration regions. 

As can be intuitively realized the acceleration of a particle with a charge $Z$ is much easier than the acceleration of a proton. These eases the conditions at the acceleration site. Particularly important is the much lower ($\propto Z^{-2}$) limit on the sources' synchrotron luminosity. Eq.  \ref{luminosity} implies a well known  drastic difference between the possible sources of nUHECRs and of pUHECRs.  
While AGNs with $L> 10^{45}$erg/s are rare and none exists at present within the GZK distance  $L_{GZK} \sim 100$Mpc, AGNs  with $L>10^{42}$erg/s are  numerous. The lower rigidity relaxes the most critical GZK problem, the lack of suitable accelerators within the GZK distance. 

\section{Propagation}
\label{propagation}

Not less drastic effects arise concerning the propagation of the nuclei from the source to Earth. The lower rigidity leads to  a much smaller Larmor radius:
\begin{equation}
R_L = 4 ~{\rm Mpc}~ E_{20} B^{-1}_{nG}  Z^{-1}_{26} \  , 
\label{RL}
\end{equation}
where $B$ is EGMF.  
The expected 
For  $R_L< \lambda$,  the maximal coherence length of the EGMF,
the  particles diffuse in Kolmogorov regime   with a mean free pass, $l$:
\begin{equation}\label{eql}
l \approx (R_L \lambda^2)^{1/3} = 1.6 ~{\rm Mpc} ~ (E_{20}/ B_{nG} Z_{26})^{1/3} \lambda^{2/3}_{Mpc}, 
\end{equation}
where the expected value of $\lambda$, is of order $0.1-10$Mpc.
The corresponding diffusion coefficient $D(E,B)$ satisfies:
\begin{equation}\label{eqD}
D\approx  0.85~{\rm Mpc}^2/{\rm My} ~ (E_{20} /Z_{26} B_{nG})^{1/3}  \lambda^{2/3}_{Mpc} . 
\end{equation}
The most energetic 
nUHECRs may have  $R_L(E) \gsim  \lambda$ and could be  in the transition region between the Kolmogorov  and the Bohm regimes. For simplicity I  assume in the following that all relevant particles are in  the Kolmogorov regime.  

The  maximal distance, $d_{max}(E) $, that an average nUHECR with energy $E$ traverses before photodisintegrating is:
\begin{equation}
d_{max} =\frac{(L_{GZK} l )^{1/2}} {\sqrt{3}}\approx   10~ {\rm Mpc} \left[ \frac{L_{GZK}(E)}{100 {\rm Mpc} } \left (\frac{ E_{20} \lambda^2_{Mpc} }{B_{nG}  Z_{26}}\right )^{1/3} \right]^{1/2} \ .
\label{dmax} 
\end{equation}
 With typical parameters $d_{max} \ll L_{GZK}$. This implies that nUHECR sources are much nearer than what was expected earlier.     Eq. \ref{dmax} is the second drastically different feature of a nUHECR  as compared with a pUHECR. 

Consider now a source at a distance $d$ from Earth that emits nUHECRs from time $T_{on}$ until $T_{off}$.
The maximal propagation time of the CRs is $t_{max} = {\rm Min}(L_{GZK}/c, T_{on})$. There is a negligible contribution from times prior to the arrival of the diffusion front \cite{FP00}, so the minimal propagation time is 
$t_{min} = {\rm Max}(d^2 /3D, T_{off})$.  For a uniform emission with a constant rate $\dot n_0$ the 
particles distribution satisfies:
\begin{equation}
n(r,t) \approx \frac { 2 \dot n_0 (t^{-1/2}_{min}-t^{-1/2}_{max})}{[8 \pi D(E,B)]^{3/2}}  \ . 
\label{diffuse}
\end{equation}
If the original spectrum of the particles at the source is $E^{-p}$  the observed spectrum will
be $E^{-(p+3\beta/2})$ where the diffusion coefficient satisfies $D \propto E^{\beta}$. A consistent solution with the observed $E^{-2.7}$ spectrum is obtained for an injection spectrum of $E^{-2.2}$ if the nUHECRs are in the Kolmogorov diffusion regime for which $\beta = 1/3$.  The spectrum might be\footnote{The observed spectrum above  the GZK energy is  steeper than -2.7, but this is attributed to the GZK suppression.} too steep  in the Bohm regime for which $\beta =1$. 
Kolmogorov diffusion  holds if   $B_{nG}\lambda_{Mpc} \gsim 5$.  Given the uncertainties in the EGMF structure this is a viable possibility.  

The CR distribution will be anisotropic with \cite{FP00}:
\begin{equation}
f(\theta, r, t) = (1 + \alpha \cos \theta) \frac{n(r,t) c }{4 \pi} \ .
\label{anisotropy}
\end{equation}
The anisotropy in the CR distribution is the flux weighted average of $d /2 t c$ so $\alpha \approx d/(6 t_{min}c)$.
For a source that is still active now: 
\begin{equation}
\alpha \approx  \frac{3 D}{  2 d c}  \approx 0.25 ~\left (\frac{ E_{20} \lambda^2_{Mpc} }{B_{nG}  Z_{26}}\right )^{1/3} \left (\frac{3.8 {\rm Mpc}}{d}\right ) .
\label{anisotropy2} 
\end{equation}
For a single source, 
given the observed isotropy of the  UHECR \cite{AugerIsotropy08,AugerIsotropy09} Eq. \ref{anisotropy2} might seem to be a problem. However, for lower energy UHECRs below the GZK energy the diffusion coefficient is much smaller and the anisotropy due to a local source will be erase. Above the GZk energy this anisotropy will be  smaller if the source has turned off at $T_{off} > d^2 /3D$. Additionally,  and more important, this (energy dependent) anisotropy will be erased by strong deflections in the Galactic magnetic fields. 
The magnetic field in the Galaxy is of order $4 \mu$G. It is composed of an ordered component and a random component. The Larmor radius in this field of even the highest energy nUHECR is  $R_{L_{Gal}} \lsim $kpc. For most arrival directions this  is smaller than the distance a nUHECRs  traverses in the Galaxy leading to a significant deflections that will erase the anisotropy of the flux reaching the Galaxy.  Numerical simulations of pUHECRs find Galactic deflections angels of a few degrees \cite{TakamiSato08}  corresponding to large ($90^o$) deflections for nUHECRs.

\section{Sources}
\label{sources}
The new limits on  nUHECR sources are quite  different from those for pUHECRs. 
(i) The lower luminosity limit (Eq. \ref{luminosity}) increases significantly the number of source candidates. (ii) On the other hand the  small maximal distance (Eq. \ref{dmax}) decreases significantly the volume where the sources can be. (iii) The larger deflections in both the EGMF  and in the Galactic magnetic field  erase most if not all  directional information on the source. 

Clearly if the EGMF is sufficiently small $B_{ng}< 0.01$,  as has been recently suggested \cite{NeronovSemikoz09,AndoKusenko10},
its effect on the nUHECRs would be minimal and there won't be a significant new lower limit on the distance of nUHECR sources. 
However, (i) and (iii) would still be applicable.
The later will be valid since nUHECRs will still suffer  strong deflections in the Galactic magnetic field. Thus even in this case  a significant correlation of the arrival direction of nUHECRs with their sources should not be expected. 

However, the common understanding is that the EGMF is stronger, of the order of a few nG.  
Upper limits on rotation measures (RM) of radio signals from distance Quasars were used \cite{Kronberg} to 
set a limit of $B<10^{-11}$G for a homogenous field and a $B<$nG for a random field with a coherence scale $\lambda_{Mpc}\approx 1$. However $\Omega_B=1$ was used  in these estimates to obtain the electrons' density. With $\Omega_B \approx 0.04$ the actual upper limits are  larger by a factor of 20, reaching $B_{nG} \approx 20$ \cite{FP00a}. More recent work suggest that the EGMF follows the large scale structure with magnetic fields of order a $\mu G$ within clusters \cite{Govoni04} and $10-100$nG within superclusters \cite{BlasiOlinto98,Ryu08}. There has been an extensive work on propagation of UHECRs in such magnetic fields (see e.g. \cite{KoteraLemoine08,Ryu10} and references therein). However, most if not all, this work is concerned with the  propagation of pUHECRs. Scaling the results  to nuclei  suggests that nUHECRs  are in the strong scattering regime. For example \cite{KoteraLemoine08} find, for a realistic magnetic field distribution that follows the large scale structure,  that the optical depth for a magnetic deflection of a pUHECR  is larger than unity and the overall deflection angles are of a few degrees. These results suggest that  the local universe is magnetically opaque for nUHECRs whose typical deflections are of order unity.

If  the  EGMF is of order of a few nG or if the magnetic field in the Virgo supercluster is of order 0.1$\mu$G then it follows from  Eq. \ref{dmax} that the  dominant source is
the nearest active radio galaxy Centaurus A  (Cen A). M87 may possibly add a minor contribution. 
At a distance of only $3.8 \pm 0.1$Mpc \cite{Rejkuba04} Cen A proves to be an excellent source candidate.  Photon emission from the nucleus of the galaxy has been detected in the radio, infra-red (IR), X-ray, and in the GeV-TeV range. Radio and GeV emission was also observed from  the large radio lobes that extend up to 250 kpc.  Cen A is also the nearest extra galactic source of  TeV photons which, apart from UHECRs, are the highest energy particles observed so far on Earth. As such they may be a good indication for production of  UHECRs as well. In the past, Cen A was already suggested as a major  (possibly only) source of UHECRs 
\cite{Cavallo78,Romero96,FP00,FP01}.   This  possibility has received a lot of attention recently following the observations of a possible 
($\sim 2\%$ significance)   
concentration of UHECR with arrival directions within a few degrees from Cen A \cite{AugerCenA}. However, if UHECRs are indeed iron nuclei this association is most likely spurious (unless both the EGMF and the Galactic magnetic fields in the direction of Cen A are particularly low). The role of Cen A as the dominant nUHECR source arises from the small maximal distance 
that nUHECR can propagate even in a modest EGMF!  Cen A is the strongest and possibly only known source within the reasonable distance. M87, at 16Mpc, might also contribute, if the EGMF is on the lower side. 

The observed present total luminosity of Cen A is  $\sim 10^{43}$erg/s, of which about half is in high energy \cite{Israel98}:  comfortably  above the synchrotron luminosity limit of Eq. \ref{luminosity}. Let $\epsilon_U$  be the efficiency of UHECR production compared
to photon production extrapolated to $10$EeV using equal power per decade and 
$\epsilon_i$ be  the fraction of energy that goes to accelerate iron nuclei.   If Cen A 
is the only source in the magnetic GZK volume of $(4 \pi/3) d_{max}^3$ then the local UHECR injection rate is
 $ \epsilon_U \epsilon_i 5.5 \times 10^{45}$ erg/Mpc$^3$/yr. This is easily consistent with the observed
value of $0.45 \pm 0.15 \times 10^{45}$ erg/Mpc$^3$/yr \cite{Katz09}, even considering the fact that the majority of the accelerated particles are lower energy protons and only a small fraction of the total UHECR energy might go to iron nuclei. 
 
Cen A is a  nearby source (in terms of $d/l$). This makes the analysis somewhat problematic in view of the  large uncertainty in the value of the intervening magnetic field and hence in the mean free path, $l$, and the corresponding diffusion coefficient. With a low value of the magnetic field the highest energy nUHECRs could be in the Bohm diffusion regime and with $d \approx l$. This would, of course make some of our simple estimates, in particular Eqs. \ref{diffuse}, \ref{anisotropy} and \ref{anisotropy2} for the nUHECR density and the anisotropy invalid. 
Another complication, which is ignored here, arises if there are large voids (of orders of tens of Mpcs) in the magnetic field structure that allow nUHECRs to propagate in straight lines over large distances. 
Detailed simulation  \cite{Isola02} suggest  isotropy for  pUHECRs  emitted from Cen A with $B\sim 1 \mu$G. This  corresponds to $50$nG 
for nUHECRs. This is comfortably within the expected estimates within the local supercluster.  Similarly, simulations of pUHECR propagation in the Galactic fields \cite{TakamiSato08} yield deflections of a few degrees, which imply defections of order unity for nUHECRs. It is interesting to note that within this model one cannot expect a complete isotropy as  both the EGMF magnetic fields and the galactic field could induce an anisotropy arising from their own structure. For example a strong EGMF in the Virgo supercluster could induce excess of particles around the super-galactic plane.

\section{Conclusion}

 {\it If UHECRs above the GZK are dominantly iron nuclei and if the EGMF is larger than a few nG}  then  the largest possible distance of the source would be only a few Mpc. Even though it is much easier to accelerate iron nuclei than protons the only available source within such a distance is 
Cen A. Large deflections of order unity in both the EGMF and in the Galactic magnetic fields would erase, in such a case,  most the directional information on the UHECR sources.


\begin{acknowledgments}
The research was  supported in part by an ERC advanced research grant and by the Center of Excellence in High Energy Astrophysics of the Israel Science Foundation. I thank Omer Bromberg, Franck Genet, 
Ehud Nakar, Elena Rossi, Re\'em Sari and Nir Shaviv for  helpful discussions and comments.
\end{acknowledgments}


\end{document}